\documentclass[aps,prd,onecolumn,amssymb]{revtex4}
\usepackage[utf8]{inputenc}
\usepackage{graphicx,bm,color}
\usepackage{amsmath}
\usepackage{amssymb}
\usepackage{amsfonts}
\usepackage{epsfig}
\newcommand{\be}{\begin{equation}}
\newcommand{\ee}{\end{equation}}
\newcommand{\bea}{\begin{eqnarray}}
\newcommand{\eea}{\end{eqnarray}}
\newcommand{\beaa}{\begin{eqnarray*}}
	\newcommand{\eeaa}{\end{eqnarray*}}

\allowdisplaybreaks[4]

\begin{document}

\title{  Quantum circuit simulation of   black hole evaporation and Maxwell 
demon interpretation }

\author{H. Hadi}
\email{hamedhadi1388@gmail.com}
\affiliation{Faculty of Physics, University of Tabriz
	Tabriz 51666-16471, Iran}

\author{Amin Rezaei Akbarieh}
\email{am.rezaei@tabrizu.ac.ir}
\affiliation{Faculty of Physics, University of Tabriz
	Tabriz 51666-16471, Iran}

	\author{Emmanuel N. Saridakis}
\email{msaridak@noa.gr}
\affiliation{National Observatory of Athens, Lofos Nymfon, 11852 Athens, Greece}
\affiliation{Departamento de Matem\'{a}ticas, Universidad Cat\'{o}lica del 
Norte, Avda.
Angamos 0610, Casilla 1280 Antofagasta, Chile}

\affiliation{CAS Key Laboratory for Researches in Galaxies and Cosmology, 
Department of
Astronomy, University of Science and Technology of China, Hefei, Anhui
230026, P.R. China}

\begin{abstract}
	
We suggest a quantum circuit model which simulates the black-hole evaporation process.
In particular, Almheiri-Marolf-Polchinski-Sully (AMPS) paradox and the ER=EPR correspondence are reconsidered
regarding our proposed model, which assumes a Maxwell's demon operating within a black hole interior.
In other words, we form a quantum circuit, mimicking the behavior of the entanglement structure of the near-horizon region and the early Hawking radiation located far from the black hole. Furthermore, we indicate how the demon, by applying nonlocal correlations, can mediate via Einstein-Rosen bridges for the purpose of simulating the transfer of quantum information across the horizon without
violating the monogamy of entanglement. Finally, the thermodynamic cost of the demon's operations regarding Landauer’s principle is
analyzed. This indicates that the information erasure has an energy comparable to the black hole entropy.

\end{abstract}
\maketitle

\section{Introduction}

The complementarity is an important concept in the physics of black holes \cite{Almheiri:2012rt}.
It claims that the entire spacetime created by a black hole cannot be fully described by either a stationary observer or an observer
falling into the black hole separately. In other words, the full information of the spacetime can be probed when
these two observers are considered simultaneously.
However, it seems that the concept of complementarity is threatened by
the existence of a firewall inside the event horizon. This feature of the horizon ultimately leads to a violation of
the equivalence principle of General Relativity.
In order to solve this problem while maintaining the principles of complementarity
and equivalence, the ER=EPR correspondence is proposed
\cite{Susskind:2014yaa}.
In this correspondence, ER is an abbreviation for the Einstein-Rosen
bridge and EPR represents the Einstein-Podolsky-Rosen pair.
It is important to emphasize that the ER=EPR correspondence, besides preserving the equivalence
principle and the complementarity of black holes, also resolves the AMPS
(Almheiri-Marolf-Polchinski-Sully) paradox \cite{Almheiri:2012rt,amps2}.

The AMPS find that accepting the complementarity of black holes and the following
three assumptions leads to a contradiction. The three assumptions are as follows:
(I) The unitary evolution of the black hole: The black hole evaporation process preserves the quantum information, and it does not destroy information.
(II) The equivalence principle: An infalling observer into the black hole does not encounter any unusual features at the event horizon.
(III) The existence of a relativistic effective quantum field theory: For a stationary external observer, the relativistic effective quantum field theory remains coherent.

Our knowledge of black holes claims that the late radiation 
$B$ of an old black hole
\cite{Page:1993df,Page:1993up}, which has emitted half of its radiation, is maximally entangled with the
early radiation $R_{B}$. This claim results from assumptions (I) and (III) and admits the entanglement of
$B$ with a subsystem of $R_{B}$ , while assumption (II) results in entanglement between 
$B$ and the
interior radiation $A$
of the black hole. Therefore, these considerations violate
the monogamy of entanglement \cite{mon1,mon2}. In this scenario, the monogamy of entanglement asserts that 
$B$ cannot be entangled with both  $A$ and $R_{B}$. To avoid this paradox, AMPS suggests that the entanglement
between $B$ and $A$ breaks down, leading to the release of a significant amount of energy at the black hole horizon.
This amount of energy is a source for creating a firewall, which extends the singularity to the horizon, leading to the absence of
spacetime within the horizon interior
\cite{Almheiri:2012rt,amps2}.

The ER=EPR correspondence \cite{cool} is a possible solution to the AMPS paradox. By applying this correspondence, one not only solves the AMPS
puzzle but also preserves the equivalence principle in the horizon region of the black hole.
The construction of the ER bridge can be explained by EPR micro-states in entangled black holes. The studies \cite{cool1, cool2}
support this interpretation. In other words, one can explain the quantum EPR correlation system as a weakly coupled Einsteinian gravity, which can clarify the ER bridge's nature. Some fundamental works claim the existence of a quantum bridge for every single quantum entangled state.

Locality is one of the controversial assumptions of AMPS. Although there is experimental evidence for locality,
one can desire the violation of it in quantum gravity in order to maintain unitarity to avoid the dramatic
consequences of firewalls and without violating our experimental evidence for locality \cite{antiamps}.
Giddings and collaborators
has notably advanced the study of unitary nonlocal models in quantum 
gravity \cite{non1,non2,non3,non4,non5,non6,non7,non8,non9,non10,non11,non12,
non13,non14,non15,non16,non17,non18}. For additional black hole solutions that 
apply nonlocality and qubit toy models to consider the physics of black hole 
evaporation, one may refer to 
\cite{nonn1,nonn2,nonn3,nonn4,nonn5,nonn6,nonn7,
nonn8,  nonn11,
nonn12,nonn13,nonn14, nonn16,nonn17,Capozziello:2012zj,
nonn19,nonn20,nonn22,nonn23,
nonn24,nonn24b,nonn25,nonn26,nonn27,nonn28,nonn29}.

In this work we investigate the nonlocality of gravitational effects in black holes. It is done by exploring
EPR pairs in a spacetime created by a black hole by using a quantum circuit model. 
Our model and studies like this may shed light on the mechanism underlying the ER=EPR correspondence. 
 Regarding this, we assume a Maxwell demon working in conjunction within a black 
hole by applying an ER bridge to study the ER=EPR correspondence.
For this purpose the demon engages in a quantum circuit.

The concept of the Maxwell's demon, which was initially introduced by James Clark Maxwell in 1867,
as a thought experiment, indicates that the second law of thermodynamics is statistical. This point of view is different from the laws
of Newton, which are based on dynamical laws. The Maxwell
demon posed a paradox, which was solved by Landauer in 1961 by applying the concept of logical irreversibility in the process of memory erasure \cite{2m}. The principle asserts that information erasure is a logically irreversible process involving energy
dissipation, which increases the entropy in the environment. Besides the significant role of Maxwell's demon in unfolding the
connections between thermodynamics \cite{3m} and information
theory, it has also been extensively studied in quantum
thermodynamics and quantum information theories \cite{4m,5m}.

The applications of Maxwell's demon in quantifying the amount of entanglement and the quantumness of
correlations can be found in \cite{10m,11m,12m}. In addition, the amount of
work that can be extracted from a correlated pair \cite{14m,15m,16m}, using Maxwell's demon, has become of interest among
researchers over the past two decades.
In parallel, some investigations focus on quantum Maxwell demons and heat engines respecting
Maxwell demon \cite{17m,18m,19m,20m,21m}.

In this paper, the entanglement between early and late Hawking radiation of an old black hole is investigated
by simulating it through a quantum circuit using a Maxwell's demon situated
inside the black hole and using the Einstein-Rosen (ER) bridge for this
purpose. We construct a quantum circuit in collaboration with a Maxwell's demon to
model this entanglement. In this approach, an observer outside the old black hole and another observer detecting early Hawking radiation
far from the black hole can identify these entangled pairs.
The paper is structured as follows: In Section \ref{Sec2},
we present a model for the Hawking radiation using a Maxwell's demon. Section
\ref{Sec3} focuses on the design of a quantum circuit that aids a Maxwell's demon in the evaporation of a black hole. Finally, we conclude with the summary section

\section{Modeling  Hawking radiation via Maxwell demon}
\label{Sec2}

 The AMPS paradox plays an important role in 
understanding   the black hole evaporation process and Hawking radiation. As we 
has mentioned in the Introduction, ER=EPR  can be a resolution for 
this paradox. In this section we suggest a quantum circuit which can play the 
role of ER bridge in the ER=EPR correspondence. Additionally, we apply it for 
an old black hole and discuss the consequences.
 
\begin{figure}[ht]
 	\centering
 	\includegraphics[width=0.9\columnwidth,angle=0]{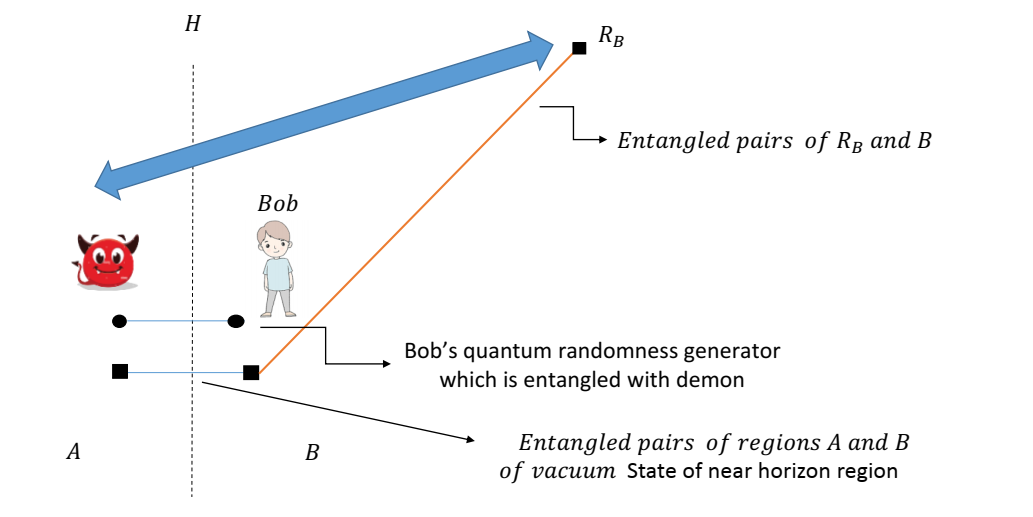}
 			\vspace{2cm}
 		\caption{%
 			\textit{%
 				Measuring of the near-horizon vacuum state and early Hawking radiation using two observers and one demon inside the black hole. 
 				$A$ and $B$ indicate the near-horizon region of inside and outside the black hole, respectively. 
 				Additionally, $R_{B}$ denotes the early Hawking radiation, and the entangled pairs between Hawking radiation with late radiation $B$ are also shown. 
 				The outside and inside of the near-horizon region are entangled, and the demon can entangle with Bob's quantum randomness generator. 
 				The two-sided arrow indicates a map between $A$ and $R_{B}$. 
 				Charlie is not depicted in the figure (see text).
 			}%
 		}
 		\label{fig1}
 	\end{figure}

 We posit the  presence of a Maxwell demon within the event horizon of a black 
hole, performing quantum computations to simulate the entangled pair between the 
near-horizon region $B$ and the early Hawking radiation $R_{B}$. In Fig. 
 \ref{fig1}  we present   an illustration of the near-horizon region 
and Hawking radiation. Before  discussing the 
details of this model, it is constructive to review the specific characteristics 
of the near-horizon region and early Hawking radiation in an old black hole.
 
 We assume that the old black hole has evaporated half of  its entropy, and 
the Hawking radiation is now far removed from the near-horizon region of the 
black hole. According to black hole physics, the region $A$ inside the black 
hole is entangled with the outside region $B$. Additionally, the states in 
region $B$ are entangled with the early Hawking radiation $R_{B}$. Hence, the 
monogamy of entanglement prevents region $B$ from being simultaneously entangled 
with both $A$ and $R_{B}$. 

 To resolve this paradox within  the ER=EPR paradigm, as illustrated in Fig. 
\ref{fig1}, the states of the Hawking radiation $R_{B}$ are mapped to the 
states of region $A$ inside the black hole as 
  \begin{equation}
  	|n \rangle_{R_{B}} \rightarrow |n \rangle_{A}, 
  \end{equation}
  through the ER bridge.  Consequently, the states of $B$ are entangled with 
$R_{B}$, and $A$ is merely a map of $R_{B}$, thereby avoiding any 
contradiction with the monogamy of entanglement. In other words, Charlie, in 
essence, can utilize a quantum computer near the Hawking radiation $R_{B}$ of an 
old black hole to compress the radiation and form a new black hole. This new 
black hole is linked to the old black hole through a wormhole 
\cite{Hawking:1987mz}. Inside the old black hole, Alice can encounter Charlie as 
he descends into his own black hole. Therefore, an observer can deduce that 
$R_{B}$ is entangled with $B$, and by entering the old black hole  can observe 
that $B$ is also entangled with $A$. However, $R_{B}$ serves as a representation 
of $A$, and there is no inconsistency.

Now we describe and simulate the  phenomenon of Hawking evaporation via a quantum circuit.
The circuit considers an entangled pair 
consisting of $B$ and $R_{B}$, equipped with a Maxwell demon placed inside the black hole.
Bob is positioned outside near the horizon of the black hole as an external observer; however, we
simplify our analysis by assuming the existence of a Maxwell demon inside the black hole,
as depicted in Fig. \ref{fig1}. Our knowledge of black holes asserts that the regions outside and inside the horizon are entangled,
however, the assumed demon can establish entanglement with Bob’s quantum 
randomness generator.The combined state of the demon and Bob’s quantum 
randomness generator is expressed as 
 \begin{equation}\label{comp}
 	|0\rangle_{D}\otimes \frac{1}{\sqrt{m}}\sum_{n=1}^{m}|n\rangle_{B}
 \end{equation}  
where $|0\rangle_{D}$ denotes the standard state of the demon, and $m$  
represents the total number of measurement configurations. As usual, the 
orthogonal states $|n\rangle$ are associated with various measurement settings.
 
Upon Bob's confirmation of  his state $|n\rangle_{B}$ and subsequent choice 
through measurement, the demon's state collapses to the corresponding state 
$|n\rangle_{D}$. Consequently, the composite state $(\ref{comp})$ transforms 
into
 \begin{equation}
 \frac{1}{\sqrt{m}}\sum_{n=1}^{m}|n\rangle_{D}|n\rangle_{B}
 \end{equation}
In addition to the entanglement between the demon and Bob's devices, the demon 
can control a qubit that becomes entangled with Bob's qubit. Consequently, at 
the beginning of the process, there are two entangled pairs between Bob and the 
demon. Following this, after Bob selects the measurement setting $X$ or $Z$, 
the demon can determine Bob's chosen measurement setting through the 
entanglement it shares with Bob's devices. Thus, after Bob's 
measurement which causes the collapse of the entangled state, the demon performs 
a measurement on its own state, and thereby identifying Bob's measurement 
setting $X$ or $Z$.
 
At the next step,  before Bob measures his qubit, the demon performs a 
measurement corresponding to Bob's chosen setting $X$ ($Z$). The demon then 
transmits its measurement results, for instance $+1$, in the $X$ ($Z$) 
direction, to Charlie, who is positioned near the early Hawking radiation 
$R_{B}$ far from the black hole, as illustrated in Fig.  \ref{fig1}. Note 
that Charlie is not depicted in the figure. The transmission of measurement 
results (Bob's measurement setting $X$ ($Z$) and the result of the collapse of 
the entangled pair, for example $+1$) can be achieved via two classical 
control-control-U gates and control-R gates (see Fig. \ref{fig2} below).

However, it is  important to recall that the demon is inside the black hole 
and thus it cannot send information to the outside. Nevertheless, if we 
restrict the model to the ER=EPR correspondence, this information can be 
transmitted through the Einstein-Rosen bridge. The ER bridge in this scenario 
can be explained in two different ways. Firstly, as previously mentioned, 
Charlie can use his quantum computer to collapse the Hawking radiation $R_{B}$ 
and create a new black hole that becomes entangled with the original black hole. 
These two black holes are linked by an ER bridge,   known as  wormhole, too. 
Secondly, every state of radiation $R_{B}$ is connected to the old black hole 
through ER bridges.

In the first scenario, Charlie can enter his black hole 
after transmitting the measurement results through the wormhole,  conduct a 
measurement, and discover that the state of the Hawking radiation reflects the 
state inside the black hole.This process can replicate the mapping between the 
internal states of the black hole and the Hawking radiation using a quantum 
circuit. Once the demon obtains Bob's measurement results using its own method, 
it can access Charlie's state through the wormhole. The demon can then rotate 
Charlie's state towards the $X$ ($Z$) direction based on the previous 
measurement result, such as $+1$. By sharing the measurement device result $X$ 
($Z$) with Charlie, who is inside his black hole, the demon can help simulate 
the entangled pair of $B$ and $R_{B}$. The procedure of assisting the demon 
inside the black hole  to simulate the entangled pair, can be described by a 
quantum circuit. The arguments presented for the first case can similarly be 
applied to the second case. However, our focus will remain on the first case, 
which involves two entangled black holes. 
 
 We close this section by  examining the physical constraints of the quantum 
demon. Each time the demon performs two measurements, one for the apparatus and 
the other to detect the collapsed state of the entangled pair. It transmits 
this information through the ER bridge, rotating Charlie's state in the same 
direction as the state inside the black hole, preparing for either a $X$ or $Z$ 
measurement. Consequently, Charlie can conduct a measurement and the state 
collapses to the expected state. During this process, the demon introduces two 
bits of information, and if it is to prepare for the next run, it must erase 
these two bits as heat into the environment. In order to consider the other 
entangled pair in the near horizon region, the demon must erase the bit of 
information to prepare for the next run, erasing one bit of information into 
heat equivalent to $kT\ln(2)$, where $T$ is the temperature of the environment. 
Since the demon performs two measurements in every run, it introduces $2kT \ln 
(2)$ of energy into the environment inside the black hole. In other words, in 
every run it distills one entanglement pair and introduces an amount of energy 
comparable to the entropy of one of the entangled pairs. By repeating this 
process for all entanglement pairs, the demon distills all entanglement pairs 
and introduces an amount of energy to the system comparable to the entropy of 
the black hole. In the following section we will consider this process as a 
quantum circuit, simulating the evaporation of the black hole  in more detail.

\section{quantum circuit of black hole evaporation process}
\label{Sec3}

\begin{figure}[ht]
	\centering
 	\includegraphics[width=0.9\columnwidth,angle=0] {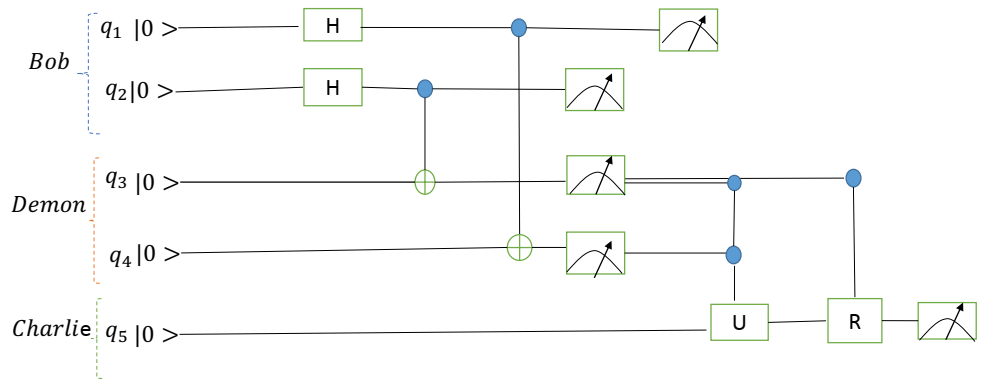}
 		\vspace{2cm}
		\caption{%
			\textit{%
				The quantum circuit: A demon inside an old black hole can 
collaborate  as part of a quantum circuit in order to simulate the  
entangled pair of Hawking radiation $B$ and $R_{B}$ of the black hole.
      }%
     }
		\label{fig2}
\end{figure}

In this section  we construct the quantum circuit for Hawking radiation and its 
measurement process, as it was illustrated  in Fig. \ref{fig1}. The 
detailed quantum circuit is depicted in Fig. \ref{fig2}.

\subsection{Quantum circuit}

A quantum  circuit implementing Maxwell demon using EPR pairs is created as 
follows. The demon is composed of two qubits, where $q_{3}$ entangles with Bob's 
random generator $q_{2}$ using a Control-Not gate to access measurement basis 
information. Additionally, $q_{4}$ acts as the random choice generator for the 
demon. Moreover, the demon is located inside a black hole and can communicate 
through a wormhole. Bob, who is outside the black hole near the horizon, also 
has two qubits. $q_{1}$ represents the qubit of the near-horizon region, 
entangled with the qubit inside the region, while $q_{2}$ serves as the random 
choice generator for his measurement basis. In this scenario, the state of 
$q_{2}$, being $|0\rangle$ or $|1\rangle$, corresponds to Pauli measurement 
basis $Z$ or $X$, respectively. As shown in the quantum circuit diagram in 
Fig. \ref{fig2}, at the start of the process, $q_{3}$ of the demon 
becomes entangled with $q_{2}$ through a Control-NOT gate, and therefore  
their state becomes
\begin{equation}
	|\psi\rangle_{23} =\frac{1}{\sqrt{2}}(|0\rangle_{2} |0 \rangle_{3} +  |1 
\rangle_{2} |1 \rangle_{3}).
\end{equation}

After Bob measures his  state $q_{2}$, causing the collapse of the state 
$|\psi\rangle_{23}$, the demon can identify $q_{3}$ as the same state as 
$q_{2}$. This implies that the demon gains immediate knowledge of Bob's basis 
information. Subsequently, the demon measures qubit $q_{4}$ before Bob measures 
his qubit $q_{1}$. Consequently, the demon knows Bob's measurement outcome 
before Bob himself. Through the wormhole, the demon can enter region $R_{B}$ and 
share the measurement outcome of $q_{3}$ with Charlie. The demon then acts on 
qubit $q_{5}$ before Charlie measures it with his chosen basis. The demon's 
objective is to rotate the state of $q_{5}$ into one of the eigenstates of the 
measurement basis. This rotation is achieved using a three-qubit classically 
Control-Control-U gate, illustrated in Fig.  \ref{fig2}.  The specific form 
of rotation $U$ is defined by
\begin{equation}
	U|q \rangle_{3}|q  \rangle_{4}|0 \rangle_{5}=|q \rangle_{3}|q 
\rangle_{4}e^{-\frac{i\pi}{4}(q_{3}+2q_{4})\sigma_{R_{Y}}}|0 \rangle_{5} ,
\end{equation} 
 where $q_{3} ~\in \{0\},~ q_{4} ~\in  \{0,1\}$ and $\sigma_{R_{Y}}$  represents 
the Pauli operator. In this case  $|q\rangle_{4}$ is $|0\rangle_{4}$ for 
$q_{4}=0$, and it is $|1\rangle_{4}$ for $q_{4}=1$. Furthermore for $q_{3}~\in 
\{1\},~ q_{4} ~\in \{0,1\}$ the rotation is given by
\begin{equation}
U|q \rangle_{3}|q \rangle_{4}|0 \rangle_{5}=|q  \rangle_{3}|q 
\rangle_{4}e^{-\frac{i\pi}{4}(q_{3}+2q_{4})\sigma_{R_{Y}}}|0 \rangle_{5} .
\end{equation} 
Note that, $|q\rangle_{4}$ is $(|0\rangle_{4}+ |1\rangle_{4})/\sqrt{2}$ for 
$q_{4}=0$, and it is $(|0\rangle_{4}-|1\rangle_{4})/\sqrt{2}$ for $q_{4}=1$.

It is important to  mention that we are exclusively focusing on the scenario 
involving two measurement settings, specifically the orthogonal Pauli 
measurement basis $Z$ and $X$. In order to incorporate additional measurement 
settings, a greater number of qubits will be necessary for the random generator. 
Furthermore, we have employed a single entangled state encompassing the $A$ and 
$B$ regions of the black hole. In order to fully explore the entire horizon 
area for Bob and the whole  radiation state $R_{B}$, a larger number 
of additional qubits will be required. The inclusion of more additional qubits 
will result in an increased need for executing multi-qubit gates within the 
quantum circuit.

\subsection{Energy dissipated by the demon }

In this subsection we examine the energy  dissipated to the system by the 
demon. With the quantum circuit model of the Maxwell demon black hole  we can 
now explore the correlation between the demon's work. Our model consists of a 
minimum demon with two qubits, $q_{3}$ and $q_{4}$, as depicted in Fig. 
\ref{fig2}. Since $q_{3}$ is entangled with $q_{2}$ and $q_{4}$ is entangled 
with $q_{1}$, there are two entangled pairs that need to be reset to the ground 
state $|0\rangle$ after each run. The possible states for qubits $q_{3}$ and 
$q_{4}$ at the end of the circuit are $|0\rangle_{3}|0\rangle_{4}$, 
$|1\rangle_{3}|0\rangle_{4}$, $|0\rangle_{3}|1\rangle_{4}$, and 
$|1\rangle_{3}|1\rangle_{4}$. Resetting the state to the ground state 
$|0\rangle$ for a qubit with the state $|1\rangle$ is equivalent to erasing $1$ 
bit of information. On average, the reset of $q_{3}$ and $q_{4}$ for $2$ 
entangled states results in the erasure of $2$ bits of information for each run. 
According to Landauer’s erasure principle, at least $2KT\ln(2)$ energy is 
dissipated into the local environment due to the work done by the demon to erase 
$2$ bits of information in each run.

Finally, note that in a more practical situation, the demon may  execute 
operations with a probability $p$ for each iteration. The simulation of the 
entangled $B$ and $R_{B}$ states, which are detected by Bob and Charlie in the 
black hole, respectively, through a Maxwell demon participating in a quantum 
circuit and operating with a probability $p$ for each iteration, is represented 
by
\begin{equation}
	\rho = p|\psi \rangle _{15} \langle \psi|+(1-p)|0\rangle_{1}\langle 
0|\otimes |0\rangle_{5}\langle 0|, 
\end{equation} 
where $|\psi \rangle_{15} $ is an entangled state  of two qubits $R_{B}$ and 
$B$.  

\subsection{  Maxwell-demon-assisting quantum circuit 
model}

It appears that  the demonstration of quantum non-locality through EPR steering 
can be replicated using local operations and classical communication. In our 
scenario, which  involves the state of near horizon $B$ and early Hawking 
radiation $R_{B}$ steering, the local operations carried out by Bob and Charlie, 
as well as the local operations of the demon and its non-local communication 
through a specific geometry, namely a wormhole, can simulate the entangled pair 
of $B$ and $R_{B}$. It is crucial to emphasize that the non-locality feature is 
made possible by the particular structure of spacetime, known as the ER bridge, 
which allows for the transmission of information through a shortcut in 
spacetime. It is worth noting that the collaboration of the demon in 
constructing a quantum circuit takes place inside the black hole, and the 
non-locality of the EPR occurs within the horizon. Additionally, the local 
operation over EPR occurs outside the horizon and can be detected by exterior 
observers. Therefore, the physics inside the black hole is responsible for the 
non-locality of the EPR, while our actions in quantum mechanics involve only 
local operations. In summary, to uncover the non-local trace of EPR and 
quantum mechanics, one must examine the other side of the horizon.

\section{Conclusions}
\label{Conclusions}
 
In this work we have constructed a quantum circuit model to simulate black 
hole evaporation, inspired by the ER=EPR correspondence and involving a Maxwell 
demon operating within the black hole interior. By combining concepts from 
quantum information theory, gravitational non-locality, and thermodynamic 
considerations, our framework reproduces the entanglement structure between the 
near-horizon modes and the early Hawking radiation. 
 
We first revisited the   
Almheiri-Marolf-Polchinski-Sully (AMPS) paradox and the ER=EPR correspondence 
within the context of this quantum circuit.
The demon, through entanglement with an external observer's quantum randomness 
generator, can access measurement information and steer external qubits via an 
ER bridge, effectively simulating the necessary correlations to maintain 
unitarity without violating the principles of monogamy.

In our  scenario, which  involves the state of near-horizon $B$ and early 
Hawking radiation $R_{B}$ steering, the local operations performed by Bob and 
Charlie, along with the local operations of the demon and its non-local 
communication through a specific geometry, namely a wormhole, can simulate the 
entangled pair of $B$ and $R_{B}$. As we discussed,  the 
non-locality feature is enabled by the particular structure of spacetime, known 
as the ER bridge, which makes possible the transmission of information through a 
shortcut in spacetime.

A key feature of our model is the explicit inclusion of the thermodynamic 
cost associated with the demon's operations. Each iteration, corresponding to 
the creation of an entangled pair between interior and exterior modes, 
necessarily involves an entropy increase consistent with Landauer's principle. 
The cumulative effect of this energy dissipation is of the order of the black 
hole's entropy, connecting the quantum circuit description to the macroscopic 
evaporation process. This connection offers a microscopic realization of how 
information flow through a wormhole could encode the Hawking radiation's 
correlations while satisfying the second law of thermodynamics.

Furthermore, our model reveals the subtleties involved in reconciling 
locality and unitarity in black hole physics. The simulated quantum steering 
between the late and early Hawking radiation - assisted by the demon and 
mediated via the wormhole geometry - reflects an effective nonlocality embedded 
in spacetime structure. Hence, from the perspective of external observers, 
only local operations and classical communication are visible, preserving 
operational locality while allowing for nonlocal correlations at the quantum 
gravitational level. This supports the view that spacetime entanglement, as 
captured in the ER=EPR conjecture, can be applied as the underlying mechanism 
ensuring consistency between quantum mechanics and gravitational physics.

In summary, the quantum circuit model we propose offers a framework 
to study aspects of black hole evaporation, quantum information flow, and 
spacetime geometry in a unified setting. Future work could extend this approach 
to more general black hole backgrounds, include more complex measurement bases, 
or incorporate noise and decoherence effects. Additionally, it would be valuable 
to further explore the relation between the demon's energy dissipation and the 
  dynamics in holographic systems. These investigations are left for future 
projects.

\section*{Acknowledgements}
This work is supported by the University of Tabriz under the title
 of Scientific Authority with contract number 133.
ENS gratefully acknowledge  the contribution of 
the LISA 
Cosmology Working Group (CosWG), as well as support from the COST Actions 
CA21136 - Addressing observational tensions in cosmology with 
systematics and fundamental physics (CosmoVerse) - CA23130,  Bridging 
high and low energies in search of quantum gravity (BridgeQG)  and CA21106 -  
 COSMIC WISPers in the Dark Universe: Theory, astrophysics and 
experiments (CosmicWISPers).


\end{document}